\documentclass[aps, prd, twocolumn, superscriptaddress, nofootinbib]{revtex4}
\usepackage{graphicx}
\usepackage{graphics}
\usepackage{dcolumn}
\usepackage{amssymb}

\begin{document}

\newcommand{\Eq}[1]{\mbox{Eq. (\ref{eqn:#1})}}
\newcommand{\Fig}[1]{\mbox{Fig. \ref{fig:#1}}}
\newcommand{\Sec}[1]{\mbox{Sec. \ref{sec:#1}}}

\newcommand{\PHI}{\phi}
\newcommand{\PhiN}{\Phi^{\mathrm{N}}}
\newcommand{\vect}[1]{\mathbf{#1}}
\newcommand{\Del}{\nabla}
\newcommand{\unit}[1]{\;\mathrm{#1}}
\newcommand{\x}{\vect{x}}
\newcommand{\ScS}{\scriptstyle}
\newcommand{\ScScS}{\scriptscriptstyle}
\newcommand{\xplus}[1]{\vect{x}\!\ScScS{+}\!\ScS\vect{#1}}
\newcommand{\xminus}[1]{\vect{x}\!\ScScS{-}\!\ScS\vect{#1}}
\newcommand{\diff}{\mathrm{d}}

\newcommand{\be}{\begin{equation}}
\newcommand{\ee}{\end{equation}}
\newcommand{\bea}{\begin{eqnarray}}
\newcommand{\eea}{\end{eqnarray}}
\newcommand{\vu}{{\mathbf u}}
\newcommand{\ve}{{\mathbf e}}


\title{Dimensional reduction in the sky}

\newcommand{\addressImperial}{Theoretical Physics, Blackett Laboratory, Imperial College, London, SW7 2BZ, United Kingdom}
\newcommand{\addressRoma}{Dipartimento di Fisica, Universit\`a “La Sapienza”
and Sez. Roma1 INFN, P.le A. Moro 2, 00185 Roma, Italia}

\author{Giovanni Amelino-Camelia}
\affiliation{\addressRoma}
\author{Michele Arzano}
\affiliation{\addressRoma}
\author{Giulia Gubitosi}
\affiliation{\addressRoma}
\author{Jo\~{a}o Magueijo}
\affiliation{\addressImperial}
\affiliation{\addressRoma}

\date{\today}

\begin{abstract}
We explore the cosmological implications of a mechanism found in several approaches to quantum-gravity, whereby the spectral dimension of spacetime runs from the standard value of 4 in the infrared (IR) to a smaller value in the ultraviolet (UV).  Specifically, we invoke the picture where the phenomenon is associated with modified dispersion relations.
With minimal assumptions, we find that UV behaviour leading to 2 spectral dimensions results in an exactly scale-invariant spectrum of vacuum scalar and tensor fluctuations, regardless of the equation of state.  The fluctuation production mechanism is analogous to the one known for varying speed of sound/light models and, unlike in inflation, the spectrum is already scale-invariant before leaving the horizon, remaining so after freeze-in. In the light of Planck's recent results we also discuss scenarios that break exact scale-invariance, such as the possibility that the spectral dimension runs down to a value slightly higher than 2, or runs down to 2 but with an extremely slow transient. We further show that the tensor to scalar ratio is fixed by the UV ratio between the speed of gravity and the speed of light. Not only does our model not require inflation, but at its most minimal it seems incompatible with it. In contrast, we find that running spectral dimensions can improve the outlook of the cyclic/ekpyrotic scenario, solving the main problems present in its simplest and most appealing realisations.
$~$\\
\end{abstract}


\maketitle


\section{Introduction}\label{intro2}
Over the past few years it has emerged that the phenomenon of running spectral dimensions is a rather generic prediction of the formalisms in use for the study
of the quantum-gravity problem.  It was first observed in computer simulations of Causal Dynamical Triangulations~\cite{Lollprl}.
It was then also found in asymptotically safe Quantum Einstein Gravity~\cite{Litim,Reuter}, Ho\v{r}ava-Lifshitz gravity~\cite{HL,HLspec},
spacetime noncommutativity \cite{Alesci:2011cg,Benedetti:2008gu}, spin foams~\cite{Modesto1,Caravelli,Magliaro,Modesto2}, multi-fractional space-times~\cite{Calcagni1,Calcagni2}
and models centered on the possibility of modified on-shell relations~\cite{visser}.  Finding observable consequences of dimensional reduction is therefore of great importance for quantum-gravity research.

In this paper we argue that running spectral dimensions are relevant in cosmology and could leave traces in the observables supplied by current cosmological data sets.
In doing so, we shall not probe running of spectral dimensions in full generality.  The present understanding of spectral dimensions is confined to a characterization in terms
of return probabilities for diffusion times in fictitious diffusion processes, a formulation way too abstract to be directly applicable to cosmology in its current state.  Instead we propose that, at least exploratively, phenomenology be derived from the observations reported in Refs.~\cite{HLspec,visser} suggesting that a large variety of scenarios for the running of the spectral dimensions can be derived from modified on-shellness/dispersion relations of the type:
\be\label{ddr1}
E^2=p^2(1+(\lambda p)^{2\gamma}).
\ee
Such modifications of the dispersion relation have long been of interest, for
independent reasons, in quantum-gravity
research~\cite{grbgac,gampul,urrutia,gacmaj,gacdsr,kowadsr,MagleePRL,leedsrPRD,rainbowDSR,jurekDSRnew}
and are here of particular interest since different choices of the parameter $\gamma$ produce some of the most studied
scenarios for running spectral dimensions. In particular, adopting (\ref{ddr1}) in a spacetime with $D+1$ Hausdorff dimensions\footnote{We refer here to the Hausdorff dimension of a Euclidean space as defined by the scaling of the volume of a ball $B(R)$ of radius $R$, e.g. in $\mathbb{R}^D$ one has $V_{B(R)}\sim R^D$ and the Hausdorff dimension is $D$.}
one finds~\cite{HLspec} that the spectral dimension for small diffusion times runs down to:
\be\label{ds}
d_S=1+\frac{D}{1+\gamma},
\ee
(in this paper we shall assume the number of spatial dimensions $D$ to be 3).  This allows us to make the important point that running spectral
dimensions can matter for cosmology by simply finding the cosmological roles of the exponent $\gamma$ and of the scale $\lambda$
determining the onset of the running.  Naturally, future studies may well show that different quantum-gravity mechanisms for achieving the same running of spectral dimensions have different implications for cosmology. Nonetheless we have established here the possible relevance of running dimensions, providing tangible phenomenological motivation for more refined model-dependent analyses.

The plan of this paper is as follows. In Section~\ref{formal} we lay out the cosmological perturbation formalism for theories with modified dispersion relations (MDR), explaining how they have the potential to produce primordial fluctuations without inflation.  In Section~\ref{rsp} we present some results on running spectral dimension from MDR.  Then, in Section~\ref{pervasive} we
highlight the observation that most quantum-gravity studies favour a value of $d_S$ of about 2, which in our approach can be modelled through values of $\gamma$ of about 2.
We make the point that running of spectral dimensions to $d_S =2$ is intimately connected with a scale-invariant spectrum of cosmological perturbations. We find that
for $\gamma=2$  the spectrum of cosmological perturbations is scale invariant, even without inflation, both for modes inside and outside the horizon, and for all equations of state.

For the rest of the paper we take this understanding for $d_S =2$ ($\gamma=2$) as the starting point for other observations of potential relevance for cosmology.
In Sections~\ref{nearsp1} and ~\ref{nearsp2} we take into account the evidence provided by Planck~\cite{Planck}, showing that the spectrum of cosmological  perturbations is indeed approximately,
but {\it not} exactly scale invariant. This leads us to suggest two possibilities. It could be that the correct UV value of $d_S$ is not exactly 2 but rather slightly larger (a possibility modelled here with a $\gamma$ slightly smaller than 2).  Alternatively we could have a {\it very} slow transient from the IR regime, with 4 spectral dimensions, to a UV regime with $d_S=2$. Intriguingly this is possible only because the CMB spectrum is slightly red, rather than slightly blue. In Section~\ref{gravwav} we compute the spectrum of primordial gravity waves. We find that the tensor to scalar ratio is controlled by the UV ratio between the speed of gravity and that of light. This would have to be rather small to comply with observations.

Our work does not require inflation or any other standard way to solve the horizon and structure formation problems.  However, in Section~\ref{othersc} we examine how our results might be combined
with inflation and the ekpyrotic Universe.  Whilst we find incompatibility with the former, we discover the pleasant result that scenarios with running spectral dimensions might fix the main
shortcomings found in the simplest realisations of the latter.  In a concluding Section we summarise our main results, highlighting the challenges they raise to the quantum gravity community.

\section{Cosmological perturbations}\label{formal}
It is known that suitable density fluctuations would be formed in the early universe as a result of a fast change in their speed of propagation~\cite{csdot}. Bimetric varying speed of light
theories~\cite{bim} and tachyacoustic cosmology~\cite{tachy} are two possible ways of putting this mechanism to work. The minimal bimetric theory leads to scale-invariant perturbations but it is also possible to obtain deviations from strict scale-invariance, subject to consistency relations involving the 3-point function~\cite{ngbim,piazza}.  It is also possible to implement this mechanism via deformed dispersion relations~\cite{Mag}, as we now review.

The class of  MDR on which we are focusing allows us a formulation in terms of a higher order derivative (HOD) theory where only higher order {\it spatial} derivatives are used, thereby fending off ghosts.
Assuming that the gravitational equations are still those of Einstein gravity, the key equation for the cosmological perturbations is then:
\be\label{veq} v''+\left[c^2 k^2 -\frac{a''}{a}\right]v=0. \ee
In terms of the variable $v$ the (comoving gauge) curvature perturbation is given by $\zeta=-v/a$.  In this equation, as usual in cosmology, modes are labeled by a comoving (constant) $k$, which is nothing but the conserved charge associated with translational invariance. However, the physical wave-number of the mode is given by:
\be
p=\frac{k}{a}
\ee
i.e. the expansion stretches the wavelength of the mode. This is the $p$ that enters the MDR, for example (\ref{ddr1}). The speed $c$ appearing in (\ref{veq}) is to be obtained from the MDR, with $c=E/p$.  Specifically, for the MDR (\ref{ddr1}) we find for $\lambda p \gg 1$:
\be\label{cofE}
c=\frac{E}{p}\propto
{\left(\frac{\lambda k}{a}\right)}^\gamma\; .
\ee
We see that if we focus on a fixed comoving mode, as is usual in cosmology, the presence of a frequency dependent speed of light translates into a time-dependent speed of light by proxy,
via the expansion\footnote{For the models considered here the phase speed ($\omega/k$) and group speed ($d\omega/dk$) are the same in the UV up to a factor of order one.}.  Considering that $a\propto\eta^\frac{1}{\epsilon-1}$ with $\epsilon=\frac{3}{2}(1+w)$, where $w=p/\rho$ is the equation of state, we can make contact with the law $c\propto \eta^{-\alpha}$ used in~\cite{csdot}, with
\be
\label{alphaeq}\alpha=\frac{\gamma} {\epsilon
-1}\; .
\ee

In (\ref{veq}) we find a competition between the pressure term (the first, describing inside-the-horizon acoustic oscillations) and the tachyonic
mass term (the second, describing the Jeans instability, for outside the horizon modes). All scenarios considered here are predicated on
having solved the  horizon problem, i.e. the first term dominates first, then the second. Since for a constant $w$ we have that $a''/a\propto 1/\eta^2$
this translates into condition $\alpha>1$ if $\eta>0$. Thus for an expanding Universe we must have:
\be
-\frac{1}{3}<w<\frac{2\gamma-1}{3}\; .
\ee
The situation is different for contracting models, as we will show in Section~\ref{ekp}.

\section{Running spectral dimension from modified dispersion relation}\label{rsp}

As mentioned in the Introduction MDR can be used as tools for realizing a running spectral dimension reproducing the dimensional reduction in the UV encountered in several quantum gravity and discrete space-time scenarios \cite{Lollprl,Litim,Reuter,HL,HLspec,Alesci:2011cg,Benedetti:2008gu,Modesto1,Caravelli,Magliaro,Modesto2,Calcagni1,Calcagni2}.  For our purposes it is useful to think of the spectral dimension as the effective dimension probed by a fictitious random walk process.  Given a (euclidean) spacetime with Hausdorff dimension $D+1$ such a process will be governed by a differential operator $\Delta_{\lambda}$, which for $\lambda=0$ reproduces the ordinary Laplacian, and a {\it diffusion time} parameter $s$.  The average return probability is given by
\begin{equation}
P(s)=\int \frac{d^{d}p \,dE}{(2\pi)^{d+1}}e^{-s\Omega_{\lambda}(p)}\, ,
\end{equation}
where $\Omega_{\lambda}(p)$ is the MDR obtained from the momentum space representation of our generalized Laplacian.  The {\it spectral dimension} is given by
\begin{equation}
d_{S}(s)= -2\frac{d\ln P(s)}{d\ln(s)}\,.
\end{equation}
Notice how $d_S$ is in general a scale dependent quantity, in particular in flat space for large values of the diffusion parameter it coincides with the topological dimension $D+1$ while in the short distance, UV, regime $s\rightarrow 0$ it is sensitive to the details of the MDR and can assume non-integer values.  It is easy to check that in a 3+1 dimensional spacetime with {\it ordinary} dispersion relation one has $d_{S}(s) = 4$ independently of the scale.  We now specialize to the case of $\Omega_{\lambda}(p)=E^{2}+f_\lambda(p^{2})$.
with
\begin{equation}
f_{\lambda}(p^{2})=p^{2}(1+(\lambda p)^{2\gamma})\, .
\end{equation}
For the case $\gamma=1$ we have (see Fig.1)
\begin{equation}
\begin{array}{lcl}
\lim_{s\rightarrow 0}d_{S}(s)&=&\frac{5}{2}\\
\lim_{s\rightarrow\infty}d_{S}(s)&=&4\,
\end{array}
\end{equation}
\begin{figure}[h]
\begin{center}
\scalebox{1}{\includegraphics{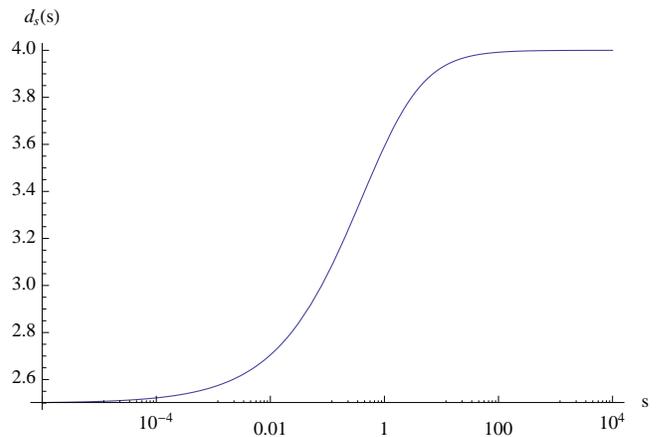}}
\caption{\label{fig:24}Running spectral dimension for quadratic plus quartic dispersion relation ($\lambda=1$)}
\end{center}
\end{figure}
For the case $\gamma=2$ (quadratic plus sextic dispersion relation) we have (see Fig.2)
\begin{equation}
\begin{array}{lcl}
\lim_{s\rightarrow 0}d_{S}(s)&=&2\\
\lim_{s\rightarrow\infty}d_{S}(s)&=&4\, .
\end{array}
\end{equation}
\begin{figure}[h]
\begin{center}
\scalebox{1.4}{\includegraphics[scale=0.7]{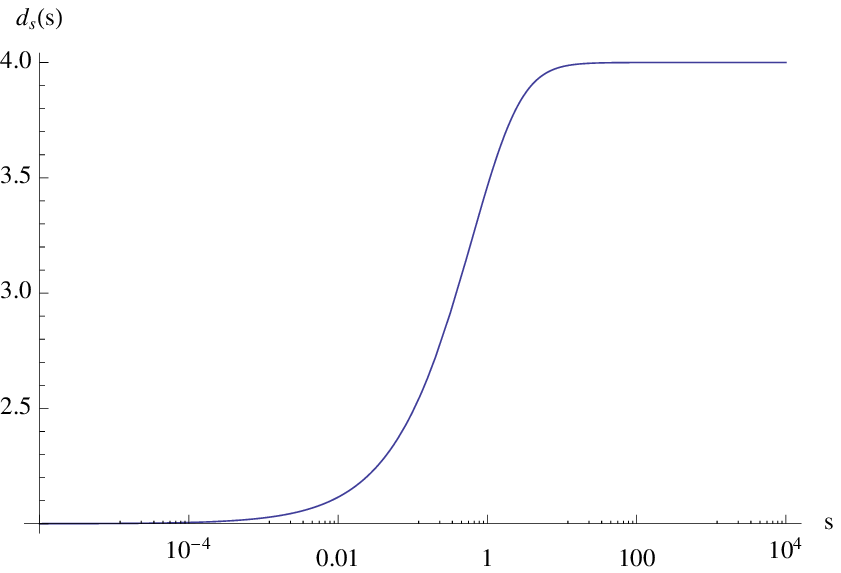}}
   \caption{\label{fig:26}Running spectral dimension for quadratic plus sextic dispersion relation ($\lambda=1$)}
\end{center}
\end{figure}

For the case $\gamma=3$ (quadratic plus eighth power dispersion relation) we have
\begin{equation}
\begin{array}{lcl}
\lim_{s\rightarrow 0}d_{S}(s)&=&\frac{4}{3}\\
\lim_{s\rightarrow\infty}d_{S}(s)&=&4 \,.
\end{array}
\end{equation}

As a last case let us consider a dispersion relation of the kind $f_{\lambda}(p^{2})=p^{2}(1+(\lambda p)^{2}+(\lambda p)^{4})$ (quadratic plus quartic plus sextic). This, as shown in Fig.3, has the same asymptotic behaviour of our case $\gamma =2$,
but with a smoother transition from the IR regime of $d_{S}(s) = 4$ to the UV regime of $d_{S}(s) = 2$.

\begin{figure}[h]
\begin{center}
\scalebox{1.4}{\includegraphics[scale=0.7]{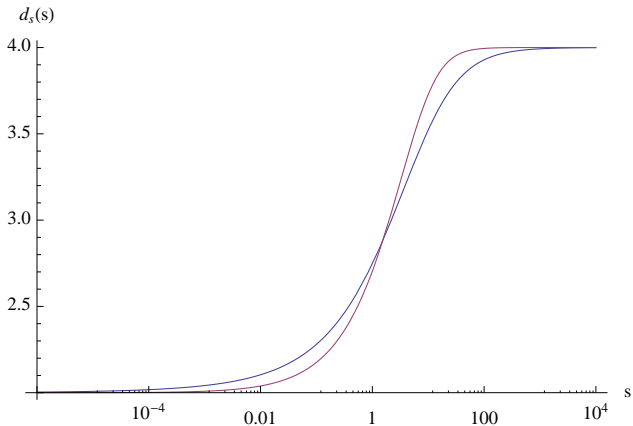}}
   \caption{\label{fig:246and26}Comparison of running spectral dimension for (blue line) quadratic plus quartic plus sextic dispersion relation ($\lambda=1$) and (purple line) quadratic plus sextic dispersion relation ($\lambda=1$). The presence of the quartic term, while leaving unchanged the asymptotic UV and IR behaviours, renders the onset of the running smoother.}
\end{center}
\end{figure}

These results show that the values of $d_{S}(s)$ in the UV and the IR are dominated, respectively, by the lower and upper value of the powers of $k$ appearing in the MDR.  In particular to the UV value of $d_{S}=2$, the one found in simulations of CDT, we find the associated MDR which {\it exactly reproduces a scale invariant spectrum} via a mechanism which we describe in the next section.


\section{Pervasiveness of  scale-invariance for $d_S=2$}\label{pervasive}

As discussed above the deformed dispersion relation
\be\label{ddr}
E^2=p^2(1+(\lambda p)^4)
\ee
appears in a number of different situations, seemingly interconnecting them. In particular it is associated with Ho\v{r}ava-Lifshitz (HL) theory for an anisotropic scaling with dynamical critical exponent $z=3$, a theory known to be power-counting  renormalisable~\cite{HL}.  Before the advent of HL theory it was known~\cite{Mag} that this dispersion relation leads to a scale-invariant spectrum of density fluctuations {\it without appealing to inflation}. This important remark was later picked up in the context of HL theory~\cite{Muko}, where it was noted that the result is independent of the assumption of detailed balance.

Here we return to the work of~\cite{Mag}, where the dependence of scale-invariance purely on (\ref{ddr}) was first studied, and investigate its significance from the standpoint of dimensional reduction.
As noted at the end of Section~\ref{formal}, for an expanding universe we need $-1/3<w<1$ for the modes to start inside the horizon. The modes can then be appropriately normalized and their vacuum fluctuations computed, and followed as the modes leave the horizon.  In~\cite{Mag} this exercise was performed, solving (\ref{veq}) in terms of Bessel functions for all values of $\gamma$ and $w$. It was found that $\gamma=2$ leads to $n_S=1$ universally. Here, instead of repeating the more rigorous treatment, we explain qualitatively the pervasiveness of scale-invariance for this dispersion relation. We use a simple trick for deriving the spectrum.  We first find out the modes' expression well inside the horizon and then their time-dependence well outside the horizon. Then we match the two expressions at horizon crossing.  This will be enough to determine the spectrum left ``frozen-in'' outside the horizon.

Using the notation $\omega = c k$, we know that inside the horizon ($\omega \eta\gg 1$) the modes are to be set to
the appropriately normalized WKB solution
\be\label{bc1}
v\sim\frac{e^{ik\int c d\eta}}{\sqrt{c k}}\sim \frac{e^{-i \beta c
k \eta}}{\sqrt{c k}} .
\ee
(where $\beta=1/(\alpha-1)>0$).
These modes will then be pushed outside the horizon ($\omega \eta\ll 1$), where we find the growing mode solution, $v = F(k) a$.
Where the two regimes meet (at $\omega\eta\sim 1$) these two solutions should match, fixing the frozen-in spectrum $F(k)$.

Why do we get pervasive scale-invariance? We see that for the dispersion relation (\ref{ddr}) inside the horizon the spectrum of $v$ is already
scale-invariant! Specifically inserting $c\approx (\lambda k/a)^2$ (c.f. Eq.~\ref{cofE})) into (\ref{bc1}) we get:
\be
v\sim \frac{ e^{-i \beta c
k \eta}}{\lambda k^{3/2}}a(\eta) \; .
\ee
Moreover, when we match this expression with $v = F(k) a$ both solutions have the same time dependence up to a phase. Therefore the spectrum left outside the
horizon is the same as inside the horizon (the factors of $a$ just cancel out when we do the matching).  This sheds light on the pervasiveness of scale-invariance for this MDR.
The background equation of state does not matter for two reasons.  The spectrum left outside the horizon is scale-invariant because inside the horizon the modes already
are scale-invariant {\it and} already have the time dependence that they will have after they leave the horizon. Therefore the details of the matching (where
the equation of state typically enters) are irrelevant: the modes simply are scale-invariant  in all regimes.

This is quite unlike inflation, where the modes start off as $v\sim e^{ik\eta}/\sqrt{2k}$ (i.e. not scale-invariant),
only to become scale-invariant {\it in a near-deSitter background} as they exit the horizon. This is because they must match $v=F(k) a$ for $k|\eta|\sim 1$,
and near de Sitter $a\sim 1/|\eta|$. This induces the extra factor of $1/k$, rendering the spectrum scale-invariant.

It is remarkable that dimensional reduction to $d_S=2$ cast in this guise (MDR, spatial HOD theory, Einstein gravity) implies strict scale-invariant fluctuations. Whilst the argument in this Section explains better why scale-invariance is so pervasive from the cosmology side, the real challenge is to understand this result directly from the properties of dimensional reduction. We will attempt this in a future publication, but this is a matter for the whole community.

\section{Further cosmological results}
In this Section we extend the previous results, which should be seen as a ``zeroth order'' solution to the problem. We explain how deviations from
exact scale-invariance might be achieved.  We also derive the spectrum of tensor fluctuations and the tensor to scalar ratio.

\subsection{Deviations from exact scale-invariance for $\gamma$ not exactly 2}\label{nearsp1}
Given the difficulties in obtaining suitable fluctuations without inflation, it is already something of a victory to derive an exactly scale-invariant spectrum without appealing to inflationary expansion.  However, the Planck results~\cite{Planck} prove conclusively that there is a small deviation from exact scale-invariance, favouring a slightly red spectrum, to a significance level of about 6 sigma. Given our results there are several possible explanations for this. At its most direct, the Planck results imply that the UV spectral dimension should be a fractional dimension slightly larger than 2.

\begin{figure}[h]
\begin{center}
\scalebox{0.7}{\includegraphics{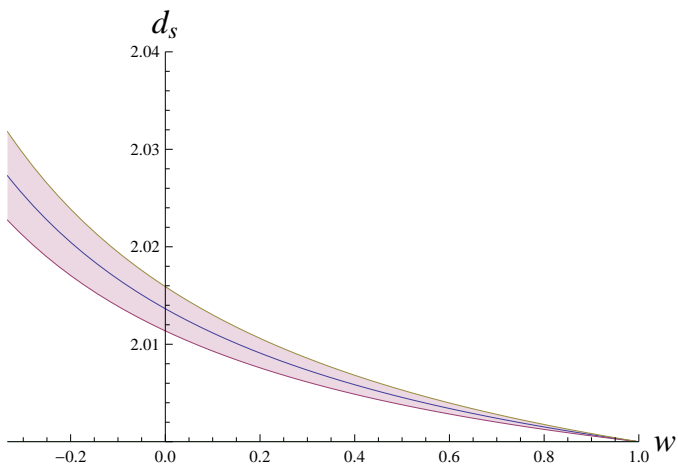}}
\caption{\label{planck}{The 1-sigma implications of the Planck results for the UV spectral dimension $d_S$, as a function of the background equation of state
$w$.}}
\end{center}
\end{figure}

It is not difficult to repeat the argument in Section~\ref{pervasive} away from strict scale-invariance ($\gamma=2$). The mode matching
procedure now gives us:
\be
F(k) a\approx \frac{1}{k^{1/2}}\left(\frac{a}{\lambda k}\right)^{\gamma/2}
\ee
to be performed at:
\be
\omega \eta =c_s k\eta\approx \left(\frac{\lambda k}{a}\right)^{\gamma} k\eta\sim 1\; .
\ee
As we see, no longer do the factors of $a$ cancel out, so the spectrum left outside the horizon is not that found inside the horizon, and the expression of $a(\eta)$ will come into play bringing the equation of state $w$ into the final result.
Carrying out the algebra we find that for a general $\gamma$ we have:
\be\label{ns}
n_S-1=\frac{\epsilon(\gamma-2)}{\gamma-\epsilon +1}\; ,
\ee
i.e. except for $\gamma=2$ and $n_S=1$, the relation between $n_S$ and $\gamma$ is $w$-dependent.  Consulting the ``dictionary'' presented in Section 2, as well (\ref{ds}), we can convert the Planck results into a band of possible $d_S$ for given $w$. The result is plotted in Fig.~\ref{planck}.

\subsection{Deviations from exact scale-invariance with slow running}\label{nearsp2}

An alternative explanation is that the in the UV $d_S=2$ but with a very long transient linking the UV and IR regimes. Therefore when the modes
we now observe first left the horizon they were experiencing an effective $\gamma$ not yet settled on 2, and so slightly smaller than 2. This
argument can only produce a slightly red spectrum, rather than blue, and even then the transient has to be cooked up to be very slow indeed, as we
now show.

Let $x=ck\eta$. For modes such that $\lambda p\ll 1$ we have $x\propto \eta\propto 1/T$ (ignoring the matter epoch), describing their evolution from
outside  ($x\gg 1$) to inside ($x\ll 1$) the horizon in the standard Big Bang Universe. Studying the evolution of the current horizon size with the temperature,
we find that when it reaches size $\lambda$ the value of $x$ is:
\be
x_*\sim {\left(\frac{T_{CMB}}{T_{Pl}}\right)}^2\frac{\lambda}{L_{Pl}}.
\ee
Including the matter epoch in the calculation slightly complicates the algebra at several steps, leading to:
\be
x_*\sim z_{eq} {\left(\frac{T_{CMB}}{T_{Pl}}\right)}^2\frac{\lambda}{L_{Pl}}
\ee
where $z_{eq}\sim 10^4$ is the redshift of matter-radiation equality.
Since the normalization of the spectrum requires $\lambda\sim 10^5 L_{Pl}$ (see~\cite{Mag} and also the next subsection),
we have that at the end of the varying-$c$ epoch in the life of this mode we must have $x_*\sim 10^{-55}$.
Before this $\lambda p\gg 1$, and so:
\be
x=ck\eta\propto c^\frac{\gamma-\epsilon+1}{\gamma}
\ee
which tells us the value of $c$ when the mode first left the horizon.

\begin{figure}[h]
\begin{center}
\scalebox{0.65}{\includegraphics{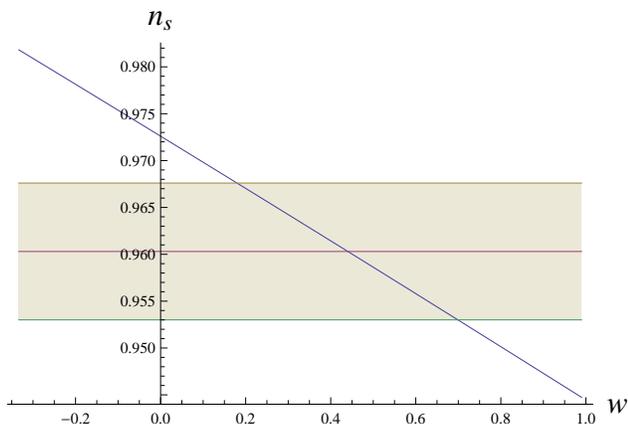}}
\caption{\label{figns} The spectral index predicted for our slow-running dispersion relation with $C=1$ as a function of the primordial background equation of state. The superposed band represents the 68\% error bar from the recent Planck results.}
\end{center}
\end{figure}

We find that
the mode that is now entering the horizon (which pegs the relevant
observational scale) first left the horizon when $c$ was:
\be
c \sim 10^{\frac{55\gamma}{\gamma+1-\epsilon}},
\ee
(using units where $c=1 $ nowadays). This is associated with momenta:
\be\label{ho}
\lambda p\sim 10^{\frac{55}{\gamma+1-\epsilon}}.
\ee
For $\gamma\approx 2$ this number is always larger than $10^{27}$, depending
on the equation of state. For $w=1/3$, we have $\epsilon=2$, and so the relevant
scale is $\lambda p\sim 10^{55}$.
Therefore if we want a dispersion relation with an effective $\gamma$ not yet settled to 2 when $\lambda p$ is this large we need a very special
expression, say of the form:
\bea
E^2&=&m^2+ p^2(1+(\lambda p)^{2\gamma(p)})\\
\gamma(p)&=&2-\frac{2}{1+C\log(1+(\lambda p)^2)}\label{gammap}\; .
\eea
This expression has the right limits ($\gamma\rightarrow 0$ for $\lambda p\rightarrow 0$; $\gamma\rightarrow 2$ for $\lambda p\rightarrow \infty$), but it exhibits a very slow UV (but not IR) transient, as required. Note the degeneracy in the UV between $C$ and the power of $\lambda p$ inside the logarithm (here chosen to be 2).
For small deviations from scale invariance  (\ref{ns}) becomes
\be
n_S-1\approx \frac{\epsilon}{3-\epsilon}(\gamma-2)
\ee
and so we can convert (\ref{ho}) (with $\gamma$ set to 2) and (\ref{gammap})
into constraint:
\be
C\approx \frac{\epsilon}{55 (1-n_S)}.
\ee
For $w=1/3$ ($\epsilon=2$) this translates into $C\approx 0.909$ (i.e.
a number of order one) in order to obtain, say $n_S=0.96$. Reciprocally,
with $C=1$ this model predicts $n_S=0.964$.
Improving the calculation (eschewing Taylor approximations) this is corrected to:
\be
n_S\approx 0.9633.
\ee
Restoring the $w$ dependence, leads to Fig.~\ref{figns}.

\begin{figure}[h]
\begin{center}
\scalebox{1}{\includegraphics{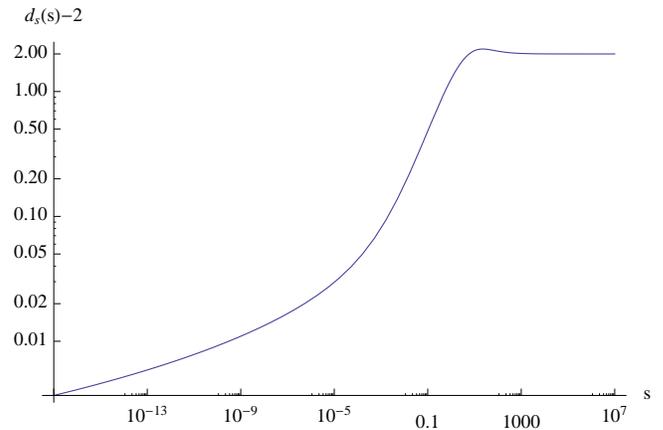}}
\caption{\label{fig:slow}Running spectral dimension for the dispersion relation exhibiting a slow UV transient, with $\lambda=1$ and $C=0.909$, such that it leads to $n_{s}=0.96$. The spectral dimension does eventually settle on 2, but with a very long transient. }
\end{center}
\end{figure}

In Fig.~\ref{fig:slow} we have plotted the running of spectral dimensions (computed as in Section~\ref{rsp}) associated with this dispersion relation.

\subsection{Generation of primordial gravity waves}\label{gravwav}
A third improvement upon our zeroth order result concerns tensor modes.
The equation for gravity waves is identical to (\ref{veq}). As a matter of fact in HL theory, strictly speaking, this equation is only valid for tensor modes, with the set up for scalar modes used in~\cite{Muko} an {\it adaptation} by analogy. Regardless of the context, it is important to point out that the MDR for gravity and for matter need not be the same, so even if the equation for the perturbations is the same, the expression for $c$ could be different.

Indeed, even if the MDR have the same form (\ref{ddr1}) and exponent $\gamma$ it could be that the scale $\lambda$ is not exactly the same, e.g. massless matter particles could satisfy (\ref{ddr1}) but gravitons, instead, be subject to:
\be\label{ddrgw}
E^2=p^2(1+b^2(\lambda p)^{2\gamma}).
\ee
The parameter $b$ is a numerical (dimensionless) factor which could arise from all sorts of reasons.
If this is the case, then we have a frequency dependent speed of light and of gravity, both as in (\ref{cofE}), but the ratio between the two at high energies is:
\be
\frac{c_{g}}{c_{m}}=b .
\ee
We now explain how $b$ is also the ratio between the amplitudes of scalar and tensor modes.  For simplicity we assume $\gamma=2$ but the calculation that follows carries
through in the more general case.

The amplitude of the scalar fluctuations spectrum may be obtained from the argument given in Section~\ref{pervasive} by keeping track of
all relevant proportionality constants. This produces result:
\be
k^3\zeta_S^2= A_S^2\sim \left(\frac{L_P}{\lambda}\right)^2
\ee
implying that $\lambda/L_P\sim 10 ^5$ (see also [14]).  Defining a variable $\zeta_T$ for tensor modes in an analogous way, the same calculation for their amplitude
leads to:
\be
k^3\zeta_T^2= A_T^2\sim b^2\left(\frac{L_P}{\lambda}\right)^2\; .
\ee
Thus the the tensor to scalar ratio is:
\be
r=\frac{A_T}{A_S}=b
\ee
(here defined in terms of variables so that $r=1$ when $c_g=c_m$).
We conclude that the high energy ratio between the speeds of light and gravity is therefore also the ratio between the amplitude of tensor and scalar modes. So as to comply with observations we therefore need the speed of gravity to be smaller than the speed of light in the UV limit.

The argument on $r$ remains unchanged when $\gamma\neq 2$, so that $n_S=n_T\neq 1$. But for all we know at this stage in the game, we could contemplate scenarios where $\gamma$ is different for matter and gravity, i.e. $\gamma_S\neq \gamma_T$, so that $n_S\neq n_T$ as well.

\section{Interaction  with other scenarios}\label{othersc}
Our model uses the mechanism for producing structure based on the varying speed of sound/light scenario~\cite{csdot,Albrecht,Moffat}. As such
it works without appealing to more conventional mechanisms, such as inflation.  Nonetheless in this Section
we examine how a direct combination of dimensional reduction
as modelled here would interact with the inflationary and ekpyrotic scenarios.

\subsection{Inflation}
At least in a direct combination, inflation and our MDR are incompatible (but we stress that a more subtle interweaving might fare better).
If $w<-1/3$, so that $\eta<0$, then the only way for the first term in (\ref{veq}) to dominate the second at early times would be to require $\alpha<1$,
i.e. $\gamma<\epsilon -1= (1+3w)/2$ (cf. Eq.~\ref{alphaeq}). Since $w<-1/3$, this implies:
\be
\gamma<0\; ,
\ee
but we already have a $\gamma=0$ term in the dispersion relations (the $E^2=p^2$ low-energy term). Thus, unless at high energies
we cancel this term (and replace it with a lower power), this cannot work, at least at face value.

\subsection{The ekpyrotic scenario}\label{ekp}
The apparent no-go we found for inflation does not apply to the ekpyrotic scenario~\cite{Khoury1,Khoury2,Stein,Gratton}.
In a contracting universe with $w>-1/3$, we have $\eta<0$. As with inflation, for the first term in (\ref{veq}) to dominate the second at early times, we still require $\alpha<1$, i.e.:
$\gamma<\epsilon -1= (1+3w)/2$ (cf. Eq.~\ref{alphaeq}). This now imposes the constraint
\be
w>\frac{2\gamma -1}{3}\; .
\ee
and there is nothing wrong with it, for a large range of $\gamma$. For our favoured (\ref{ddr}), for example, we have $\gamma=2$ and this requirement simply means $w>1$.

Remarkably, the calculation for the fluctuations described above still applies with minimal adaptation. In particular (\ref{ddr}) is
still associated with a scale-invariant spectrum (for all $w>1$, now).  The modes inside the horizon still satisfy (\ref{bc1}) and so already are scale-invariant, and proportional to $a$ (which is now decreasing in time). Outside the horizon we must check the form of the growing mode, since usually growing and decaying modes are interchanged in contracting Universes (compared to expanding ones). We find two solutions:
\bea
v_1&\propto&a\\
v_2&\propto&\frac{|\eta|}{a}
\eea
and we see that when $w>1$ the growing mode still is the mode proportional to $a$ (something that does not happen if $w<1$). As a result the mode matching exercise leads to the same scale-invariant result.

\begin{figure}[h]
\begin{center}
\scalebox{0.7}{\includegraphics{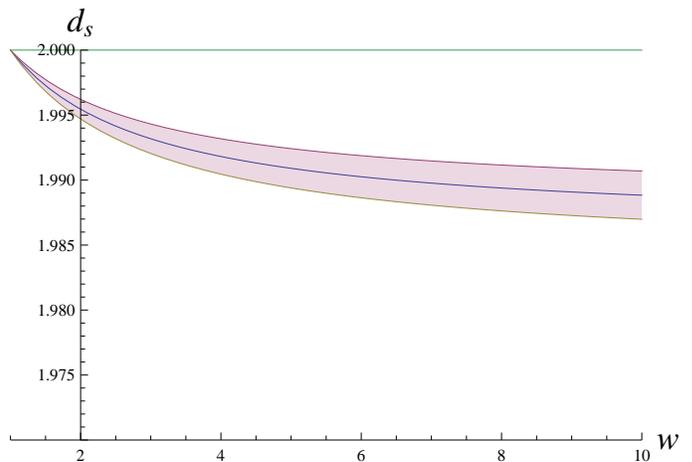}}
\caption{\label{planckekp}{The 1-sigma implications of the Planck results for the UV spectral dimension $d_S$, within a basic one field ekpyrotic model.}}
\end{center}
\end{figure}

More generally formula (\ref{ns}) is still valid in ekpyrotic scenarios when $\gamma\neq 2$, but now a red spectrum implies running to a spectral dimension {\it smaller} than 2.  The implications of Planck results in this case are plotted in Fig.~\ref{planckekp}. For large $w$ the prediction for the spectral index becomes weakly dependent on $w$, with result:
\be
n_S-1\approx 2-\gamma=3\frac{d_S-2}{d_S-1}
\ee

It is interesting that a combination of dimensional reduction (or more precisely MDR) and the ekpyrotic scenario removes some of the major shortcomings
found in its simplest and most appealing realisations~\cite{Khoury1,Gratton}.  Specifically:
\begin{itemize}
 \item We now find a scale-invariant spectrum in $\zeta$ (rather than in
$\Phi$ only).
\item The growing mode is the frozen-in time independent mode, instead of
an unstable, time-dependent mode, as in conventional scenarios.
\end{itemize}
These problems have led to the construction of more complex multifield models~\cite{Buchbinder,Lehners,Creminelli,Koyama}.
We point out here that the simplest models can be fixed, should dimensional reduction play a role close to the bounce. Furthermore we only need $w>1$, rather than $w\rightarrow \infty$, to realise
a viable structure formation scenario.


\section{Conclusions}

We conclude by laying down a series of challenges set by the results we have derived. In this paper we have shown how under basic assumptions dimensional reduction has clear implications in cosmology. The assumptions are: representation of the phenomenon by modified dispersion relations; representation of MDR by a spatial HOD field theories; and Einstein gravity. It is reasonable to expect that some of our findings will prove to be generic for theories with running spectral dimensions, rather than specific to their MDR interpretation. Still it would be important for this research programme to find more direct evidence of this general applicability.

For example, in Section~\ref{pervasive}  we showed how running to a UV 2 dimensional world is intimately associated with ``pervasive'' scale-invariance: fluctuations which are scale-invariant inside and outside the horizon, and for all equations of state. We shed light on this phenomenon from the cosmology side. However, we feel this raises another challenge directed to the quantum gravity community: to understand this remarkable result {\it directly} from the properties of spectral dimensional reduction. Such an understanding would prove the generality of our conclusion.

We then took the results on scale invariance in Section~\ref{pervasive}
as a benchmark, or a ``zeroth'' order
cosmological requirement, and went beyond models that produce strict
scale-invariance.  We found two ways in which dimensional reduction may be
tuned to produce deviations from strict scale-invariance, as those
seen by the Planck satellite
(Sections~\ref{nearsp1} and \ref{nearsp2}).
Firstly it could be that the UV spectral dimension to which
one runs is not exactly 2, but it is slightly higher. The exact
figure is not universal and depends on the background equation of state
(see Fig.~\ref{planck}). We note with interest that there is some cursory
evidence for this scenario in CDT computer simulations (e.g.~\cite{sotir}).
But it could also be
that we do run to $d_S=2$ but with a very slow UV transient. We quantified
this ``slowness'' carefully at the end of Section~\ref{nearsp2}, and
we propose that a thorough investigation of this possibility be performed in the various scenarios being considered in the
study of the quantum-gravity problem.

Beyond the ``zeroth order'' cosmology we also examined the
spectrum of gravity waves.
A further challenge to the quantum gravity community arises from the
phenomenology of tensor modes. We obtained the ratio between tensors
and scalars in terms of the UV ratio between the speed of gravity and that
of light, $b$. But for all we know from quantum gravity theory it could be
that the exponent $\gamma$ entering their MDR is different as well,
so that the tensor and scalar spectral indices are different
($n_S\neq n_T$). Currently this remains a completely unconstrained
possibility within quantum gravity theory, with no relation between
$r$, $n_S$ and $n_T$. Therefore there are no ``a priori'' consistency
conditions in terms of these observables, just like there are in
inflation, something that should be seen as a shortcoming. The possibility
that further theoretical work in quantum gravity might change this state
of affairs should therefore be cherished, and taken up as a challenge.
Ideally the theory should provide exact values for $\gamma_S$, $\gamma_T$
and $b$, but even a constraint between them would be helpful.

In this paper we restricted ourselves to issues that could be investigated
without a detailed description of interactions. This is justified
by the rather preliminary state of progress in the description
of interactions within most theories that can motivate MDRs.
However, the pay off for facing this challenge would be substantial on
the cosmology side, since it would allow one to examine
non-Gaussianities in this class of models.

We also feel that it would be important to look deeper into the interplay
between dynamical dimensional reduction and the inflationary and ekpyrotic
scenarios, which was here only preliminarily explored.

\section{Acknowledgments}
GAC, MA, and GG were supported in part by the John Templeton Foundation.  The work of MA was also supported by the EU Marie Curie Actions through
a Career Integration Grant.
JM was funded by STFC through a consolidated grant and by an International Exchange Grant from the Royal Society.

\bibliography{refsDRITS}

\end{document}